\documentclass[apl, twocolumn,superscriptaddress,tightenlines]{revtex4-1}

\usepackage{amsthm}
\usepackage{amsmath}
\usepackage{bbm, dsfont}
\usepackage{graphicx}
\usepackage[usenames,dvipsnames]{color}
\usepackage[colorlinks=true,citecolor=magenta,urlcolor=blue]{hyperref}
\usepackage[table]{xcolor}
\usepackage{colortbl}
\usepackage{color}
\usepackage{multirow}
\usepackage{hhline}
\usepackage{tabu}
\usepackage{epstopdf}
\usepackage{booktabs}
\usepackage[normalem]{ulem}

\usepackage[textwidth=3cm, textsize=footnotesize]{todonotes}

\newcolumntype{C}[1]{ >{ \centering\arraybackslash}p{#1}}
\def \minipagesize {4.1cm}

\newcommand{\dt}[3]{\cellcolor{#1} #2}

\newcommand{\tn}[1]{\textnormal{#1}}
\newcommand{\be}{\begin{equation}}
\newcommand{\ee}{\end{equation}}

\newcommand{\epscorr}{\eps_{\tn{cor}}}

\newcommand{\esec}{\eps_{\tn{sec}}}

\newcommand{\del}[1]{}

\renewcommand\tablename{Tab.}
\renewcommand\figurename{Fig.}

\newcommand{\sket}[1]{{\ensuremath{\lvert#1\rangle}}}
\newcommand{\lket}[1]{{\ensuremath{\left\lvert#1\right\rangle}}}
\newcommand{\ket}[1]{\if@display\lket{#1}\else\sket{#1}\fi}

\newcommand{\sbra}[1]{{\ensuremath{\langle#1\rvert}}}
\newcommand{\lbra}[1]{{\ensuremath{\left\langle#1\right\rvert}}}
\newcommand{\bra}[1]{\if@display\lbra{#1}\else\sbra{#1}\fi}

\newcommand{\sbraket}[2]{{\ensuremath{\langle#1\rvert#2\rangle}}}
\newcommand{\lbraket}[2]{{\ensuremath{\left\langle#1\!\left\rvert\vphantom{#1}#2\right.\!\right\rangle}}}
\newcommand{\braket}[2]{\if@display\lbraket{#1}{#2}\else\sbraket{#1}{#2}\fi}

\newcommand{\sketbra}[2]{{\ensuremath{\lvert #1\rangle\!\langle #2\rvert}}}
\newcommand{\lketbra}[2]{{\ensuremath{\left\lvert #1\right\rangle\!\!\left\langle #2\right\rvert}}}
\newcommand{\ketbra}[2]{\if@display\lketbra{#1}{#2}\else\sketbra{#1}{#2}\fi}

\newcommand{\eps}{\varepsilon}

\definecolor{gray4}{gray}{0.8}
\definecolor{gray2}{gray}{0.6}

\begin{document}
\title{Detector-device-independent QKD: security analysis and fast implementation}

\author{Alberto Boaron}
\affiliation{Group of Applied Physics, University of Geneva, Chemin de Pinchat 22, CH-1211 Geneva 4, Switzerland}
\author{Boris Korzh}
\affiliation{Group of Applied Physics, University of Geneva, Chemin de Pinchat 22, CH-1211 Geneva 4, Switzerland}
\author{Raphael Houlmann}
\affiliation{Group of Applied Physics, University of Geneva, Chemin de Pinchat 22, CH-1211 Geneva 4, Switzerland}
\affiliation{ID Quantique SA, 3 Ch. de la Marbrerie, CH-1227 Carouge, Switzerland}
\author{Gianluca Boso}
\affiliation{Group of Applied Physics, University of Geneva, Chemin de Pinchat 22, CH-1211 Geneva 4, Switzerland}
\author{Charles Ci Wen Lim}
\affiliation{Quantum Information Science Group, Computational Sciences and Engineering Division, Oak Ridge National Laboratory, Oak Ridge, TN 37831-6418, US}
\author{Anthony Martin}\email{Anthony.Martin@unige.ch}
\affiliation{Group of Applied Physics, University of Geneva, Chemin de Pinchat 22, CH-1211 Geneva 4, Switzerland}
\author{Hugo Zbinden}
\affiliation{Group of Applied Physics, University of Geneva, Chemin de Pinchat 22, CH-1211 Geneva 4, Switzerland}

\begin{abstract}
One of the most pressing issues in quantum key distribution (QKD) is the problem of detector side-channel attacks. To overcome this problem, researchers proposed an elegant ``time-reversal'' QKD protocol called measurement-device-independent QKD (MDI-QKD), which is based on time-reversed entanglement swapping. However, MDI-QKD is more challenging to implement than standard point-to-point QKD. Recently, an intermediary QKD protocol called detector-device-independent QKD (DDI-QKD) has been proposed to overcome the drawbacks of MDI-QKD, with the hope that it would eventually lead to a more efficient detector side-channel-free QKD system. Here, we analyze the security of DDI-QKD and elucidate its security assumptions. We find that DDI-QKD is not equivalent to MDI-QKD, but its security can be demonstrated with reasonable assumptions. On the more practical side, we consider the feasibility of DDI-QKD and present a fast experimental demonstration (clocked at 625~MHz), capable of secret key exchange up to more than 90~km.
\end{abstract}

\maketitle

\section{Introduction}

Secure communication is a cornerstone of our society and finding a way to protect our personal data while making it globally accessible is a profound challenge. Quantum key distribution (QKD) enables the secure establishment of cryptographic keys between two remote users, Alice and Bob~\cite{Bennett1984}.~Importantly, the security of QKD depends only on the principles of quantum physics and can be proven to be secure against quantum eavesdroppers under certain assumptions about the involved devices~\cite{Gisin2002, Dusek2006,Scarani2009}.~However, in practice, actual devices may deviate from their ideal specifications and lead to security loopholes~\cite{Lo2014}. 

In the last decade, much attention has been devoted to understanding the impact on QKD security due to the behaviour of single-photon detectors and how one can break the security of QKD by exploiting the physics of their operation.~It turns out that there are several ways to exploit the imperfections of the detectors~\cite{Zhao2008,Lydersen2010, Wiechers2011}.~These findings exemplify the fact that, like all crypto-systems, QKD is only as strong as its weakest link, despite the fact that QKD is in principle secure against general attacks.~To overcome this security loophole, researchers proposed an elegant ``time-reversal'' protocol called measurement-device-independent QKD (MDI-QKD), which is based on the principle of entanglement swapping~\cite{Lo2012,Li2014,Valivarthi2015,Comandar2015,Wang2015,Tang2016}.~More specifically, the central idea is to perform a Bell state measurement (BSM) between two qubit states, which are randomly prepared by Alice and Bob, as in the standard Bennett-Brassard 1984 (BB84) QKD protocol~\cite{Bennett1984}.~In this case, the measurement unit is seen as part of the \emph{untrusted} quantum channel and security is automatically guaranteed against all detector side-channel attacks. 

MDI-QKD is however more challenging to implement than standard point-to-point (PtP) QKD. First, it requires the interference of two independent and indistinguishable photons over long distances. This could be challenging because the photons have to simultaneously arrive at the BSM while maintaining their indistinguishability in all degrees-of-freedom (DoFs). Second, the secret key rate (SKR) is limited by the achievable coincidence rate at the BSM, which is at most 50\% of the photon detection rate assuming linear optics. Third, the finite-key analysis of MDI-QKD is less efficient than standard PtP QKD in that it requires a much larger post-processing block size~\cite{Curty2014} than its PtP counterpart~\cite{Lim2014a}.~Nevertheless, we note that an exchange of around 5~kbps over 100~km has been recently demonstrated (neglecting the finite-key effects)~\cite{Comandar2015}.

Recently, a family of QKD protocols was proposed to simplify MDI-QKD, which we collectively refer to as detector-device-independent QKD (DDI-QKD)~\cite{Lim2014, Gonzalez2015, Liang2015}. These QKD protocols use the fact that one can encode multiple qubits (using different DoFs) onto a single photon and that these qubits can be manipulated independently.~In this way, one can imagine MDI-QKD being carried out using only one photon as a carrier for Alice and Bob's qubits:~Alice first encodes her qubit into the photon and then sends it to Bob, who encodes his qubit onto another DoF. The resulting two-qubit photon is then sent to a BSM apparatus.~Therefore, only a single photon detection is required (like in PtP QKD). Since the guiding principles of DDI-QKD are similar to those of MDI-QKD, it is conjectured that the security level of DDI-QKD is comparable to MDI-QKD.~Here, we present a thorough security analysis of DDI-QKD together with a new experimental implementation of a complete crypto-system.

\begin{figure*}[t]
\includegraphics[width = 1.9 \columnwidth]{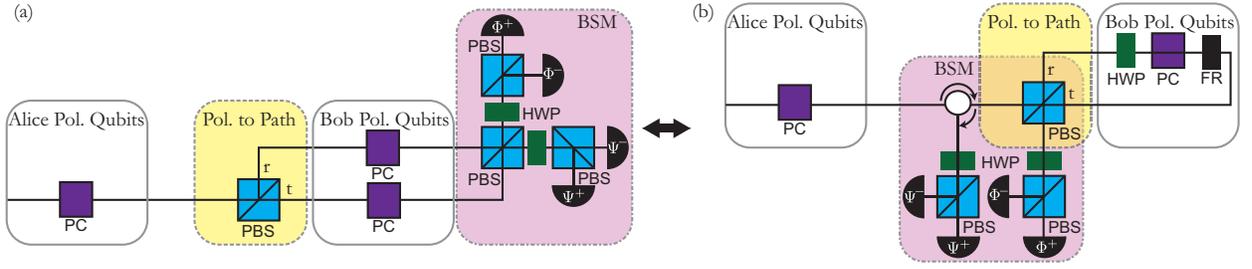}
\caption{\label{fig_prin_DDI}Conceptual setup. The schematic (b) corresponds to an alternative version of schematic (a), which is easier to implement. PC: polarization controller; PBS: polarizing beam splitter; HWP: half-wave plate; BSM: Bell state measurement; FR: Faraday rotator.}
\end{figure*}

\section{Principles of DDI-QKD}

The conceptual setup is presented in \figurename{~\ref{fig_prin_DDI} (a)}. Alice encodes a qubit $\ket{\psi_A}= \alpha_A \ket{\tilde{H}} + \beta_A \ket{\tilde{V}}$ in the polarization DoF of a single-photon and sends it to Bob. At the input of Bob a polarizing beam splitter (PBS) converts the polarization modes into spatial modes such that the qubit of Alice is converted to a state of the form $\ket{\psi_A} = \alpha_A \ket{r} + \beta_A \ket{t}$, where $r$ and $t$ represent the transmitted and reflected path of the PBS, respectively. Then, Bob encodes a qubit $\ket{\psi_B} = \alpha_B \ket{H} + \beta_B \ket{V}$ in the polarization DoF of the photon. The same polarization state needs to be encoded in the two paths. The state of the photon is then $\ket{\psi_A}\otimes \ket{\psi_B}$.

A BSM is performed by recombining the two spatial modes via a PBS and applying a projection in the basis $\left\{ \ket{+};\ket{-}\right\}$ on both output arms using two additional PBSs. $\ket +$ and $ \ket - $ correspond to $(\ket H+ \ket V)/{\sqrt{2}}$ and $(\ket H- \ket V)/{\sqrt{2}}$, respectively. A click in one of the four outputs corresponds to a projection into one of the following Bell states:
\begin{equation}
\begin{cases}
\ket{\Phi^\pm} = \frac{1}{\sqrt{2}}\left[\ket{r}\ket{H} \pm \ket t \ket V \right]\\
\ket{\Psi^\pm} = \frac{1}{\sqrt{2}}\left[\ket{r}\ket{V} \pm \ket t \ket H \right].
\end{cases}
\end{equation}

In order to exchange secret keys, the protocol is the following: Alice and Bob independently encode states randomly chosen out of the four following BB84 states $\left(\ket{H}; \ket{V}; \ket{+}; \ket - \right)$. The probabilities for each Bell state are given in \tablename{\,\ref{tab_truthtable}}. After sifting, one cannot determine the bit sent by Alice only from the knowledge of which detector has clicked. Both the result of the BSM and the state encoded by Bob are necessary to retrieve the bit chosen by Alice, using~\tablename{\,\ref{tab_truth}}.~
Before describing the practical implementation, we will take a closer look at the security.

\begin{table}[h]
\begin{tabular}{*{2}{c}}
\begin{minipage}{\minipagesize}
\begin{tabular}{C{0.4cm}*{2}{C{0.8cm}}!{\color{black}\vrule}*{2}{C{0.8cm}}}
 \arrayrulecolor{white}
 a) & \multicolumn{4}{c}{$\ket{\Phi^+}$} \\ 
 & H &  V & $+$ & $-$\\
H &\dt{gray2}{0.50}{2}&\dt{white}{0.00}{1}&  \dt{gray4}{0.25}{1}  &  \dt{gray4}{0.25}{1}\\ \hline 
 \arrayrulecolor{black}
V & \dt{white}{0.00}{0} &\dt{gray2}{0.50}{2} &  \dt{gray4}{0.25}{1} &  \dt{gray4}{0.25}{1}\\ \hline 
 \arrayrulecolor{white}
$+$ & \dt{gray4}{0.25}{2} &  \dt{gray4}{0.25}{2} &\dt{gray2}{0.50}{2} &  \dt{White}{0.00}{0}  \\ \hline
$-$ & \dt{gray4}{0.25}{1} &  \dt{gray4}{0.25}{1} & \dt{White}{0.00}{0} & \dt{gray2}{0.50}{2}  \\ \hline
\end{tabular}
\end{minipage} &
\begin{minipage}{\minipagesize}
\begin{tabular}{C{0.4cm}*{2}{C{0.8cm}}!{\color{black}\vrule}*{2}{C{0.8cm}}}
 \arrayrulecolor{white}
 b) & \multicolumn{4}{c}{$\ket{\Phi^-}$} \\ 
 & H &  V & $+$ & $-$\\
H & \dt{gray2}{0.50}{3} & \dt{white}{0.00}{1} &  \dt{gray4}{0.25}{2} &  \dt{gray4}{0.25}{2}\\ \hline
 \arrayrulecolor{black}
V & \dt{white}{0.00}{0} &\dt{gray2}{0.50}{2} &  \dt{gray4}{0.25}{1} &  \dt{gray4}{0.25}{1} \\ \hline
 \arrayrulecolor{white}
$+$ & \dt{gray4}{0.25}{2} &  \dt{gray4}{0.25}{1} &  \dt{white}{0.00}{0} & \dt{gray2}{0.50}{2}   \\ \hline
$-$ & \dt{gray4}{0.25}{2} &  \dt{gray4}{0.25}{2} & \dt{gray2}{0.50}{2}  &  \dt{white}{0.00}{0} \\ \hline
\end{tabular}
\end{minipage}  \\
\\

\begin{minipage}{\minipagesize}
\begin{tabular}{C{0.4cm}*{2}{C{0.8cm}}!{\color{black}\vrule}*{2}{C{0.8cm}}}
 \arrayrulecolor{white}
 c) & \multicolumn{4}{c}{$\ket{\Psi^+}$} \\ 
 & H &  V & $+$ & $-$\\
H& \dt{white}{0.00}{0} & \dt{gray2}{0.50}{2}  &  \dt{gray4}{0.25}{2} &  \dt{gray4}{0.25}{1}\\ \hline
 \arrayrulecolor{black}
V &  \dt{gray2}{0.50}{2} & \dt{white}{0.00}{1} &  \dt{gray4}{0.25}{1} &  \dt{gray4}{0.25}{1} \\ \hline
 \arrayrulecolor{white}
$+$ & \dt{gray4}{0.25}{2} &  \dt{gray4}{0.25}{2} &\dt{gray2}{0.50}{2} &  \dt{White}{0.00}{0}  \\ \hline
$-$ & \dt{gray4}{0.25}{1} &  \dt{gray4}{0.25}{1} & \dt{White}{0.00}{0} & \dt{gray2}{0.50}{2}  \\ \hline
\end{tabular}
\end{minipage} &
\begin{minipage}{\minipagesize}
\begin{tabular}{C{0.4cm}*{2}{C{0.8cm}}!{\color{black}\vrule}*{2}{C{0.8cm}}}
 \arrayrulecolor{white}
 d) & \multicolumn{4}{c}{$\ket{\Psi^-}$} \\ 
 & H &  V & $+$ & $-$\\
H& \dt{white}{0.00}{0} & \dt{gray2}{0.50}{2}  &  \dt{gray4}{0.25}{2} &  \dt{gray4}{0.25}{1}\\ \hline
 \arrayrulecolor{black}
V &  \dt{gray2}{0.50}{2} & \dt{white}{0.00}{1} &  \dt{gray4}{0.25}{1} &  \dt{gray4}{0.25}{1} \\ \hline
 \arrayrulecolor{white}
$+$ & \dt{gray4}{0.25}{2} &  \dt{gray4}{0.25}{1} &  \dt{white}{0.00}{0} & \dt{gray2}{0.50}{2}   \\ \hline
$-$ & \dt{gray4}{0.25}{2} &  \dt{gray4}{0.25}{2} & \dt{gray2}{0.50}{2}  &  \dt{white}{0.00}{0} \\ \hline
\end{tabular}
\end{minipage}\\
\end{tabular}
\caption{Bell state probabilities of the DDI-QKD protocol. Probabilities to measure the photon in each Bell state as a function of the qubits encoded by Alice (rows) and Bob (columns). \label{tab_truthtable}}
\end{table}

\begin{table}[h] 
\begin{tabular}{C{1.2cm}!\vrule C{1.2cm}  C{1.2cm}  C{1.2cm}  C{1.2cm}  }
Bob & \ket{\Phi^+} & \ket{\Phi^-}  & \ket{\Psi^+}  & \ket{\Psi^-} \\
\arrayrulecolor{black} \hline
H & 0 & 0 & 1 & 1\\
V & 1 & 1 & 0 & 0\\ 
+ & 0 & 1 & 0 & 1\\
$-$  & 1 & 0 & 1 & 0\\
\end{tabular}
\caption{\label{tab_truth}Truth table used by Bob to extract the bit values. For Alice, the bit values are 0 and 1 for the states $\{H; +\}$ and $\{V; -\}$, respectively. Note that the bit values of Bob depend on the qubit state and on the BSM result.}
\end{table}

\section{Security of DDI-QKD}
\label{sec.security}

The security of DDI-QKD is based on the following assumptions: i) Alice and Bob's random number generators as well as the classical post-processing are trusted. This basic assumption is necessary for all QKD schemes, including device-independent (DI-QKD) protocols. ii) Alice and Bob's linear optical circuits are fully characterized and cannot be influenced by any eavesdropper, commonly denoted as Eve. iii) Eve may exploit imperfect detectors via the optical fiber, but she has no physical access to the detectors, in particular she has no access to the outputs of the interferometer. iv) The detectors may have some defects, but are not from a malicious provider. This means they are independent of Eve. 

In the case of single-qubit quantum channels (i.e. Eve is restricted to sending pulses of light to Bob, which are on the single photon level) the first two assumptions are sufficient in order to prove formally the security of DDI-QKD. This has been shown in Refs.~\cite{Lim2014,Gonzalez2015}, and is detailed in the Appendix~\ref{sec_skr}. This also means that for this scenario DDI-QKD and MDI-QKD are equivalent.

The situation is more complicated if we consider attacks based on multi-photon states. With strong pulses, Eve could easily make a Trojan horse attack and measure Bob's settings if she had access to the output of Bob's interferometer. But this is in contradiction with assumption iii). Eve could also try a more subtle Trojan horse attack as proposed in Ref.~\cite{Qi2015}, where the detectors have shared randomness with Eve, which is in contradiction with assumption iv).~Another attack could be the siphoning attack presented in Appendix~\ref{sec_siphoning}, which works even if the quantum channel is restricted to a single spatial-temporal mode.~However, this attack is not compatible with assumptions iii) and iv). Finally, it is important to note that Trojan horse attacks based on back reflection, which can affect both Alice and Bob, have to be avoided by using a set of isolators and frequency filters as is the case for MDI-QKD as well.

Let us now consider the class of attacks based on detector blinding~\cite{Lydersen2010}. To perform such attacks, the eavesdropper shines strong classical light onto the detectors, such that they all cease to work in the Geiger mode and instead begin to operate in the linear regime. In this regime, if any of the detectors receive a light pulse which exceeds a certain threshold, a detection signal can be generated. From the perspective of Bob, this signal is indistinguishable from that generated by a single-photon detection in the Geiger mode. Let us denote by $\mu_i$ the threshold of the detector $D_i$. If only one detector (one Bell state) was used, the system is equivalent to a normal BB84 protocol and is potentially vulnerable as shown in Ref.~\cite{Lydersen2010}. However, if we consider the DDI-QKD setup with a complete BSM, such an attack will be detected by looking at the detection statistics. In the case where the thresholds of the four detectors are identical, the blinding attack will generate double detections. When this happens, Bob assigns a detection to a random detector, which affects directly the quantum bit error rate (QBER). In the case where the thresholds are different for every detector, for example if $\mu_1 < \mu_2$, then it is indeed possible to generate a detection in $D_1$ while $D_2$ does not click. However, Eve will not be able to make $D_2$ click independently. More generally, with such an attack, she will not be able to reproduce the expected detection probabilities for all detectors and all settings of Alice and Bob, as detailed in~\tablename{\,\ref{tab_truthtable}}. Note that, active randomization of the detection statistics has been proposed as a countermeasure against blinding attacks~\cite{ferreira2015safeguarding, Lim2015}. 

In short, despite their conceptual similarities, DDI-QKD is not equivalent to MDI-QKD and the additional, arguably very reasonable, assumptions iii) and iv) have to be made in order to guarantee its security. \\

\section{Experimental setup}

In our previous proof-of-principle experiment~\cite{Lim2014}, Alice and Bob used the polarization and spatial DoFs, respectively, to encode their qubits. It is challenging to achieve high encoding rates with such an implementation, because Bob's phase modulator has to be polarization insensitive, something that is not possible with high-speed electro-optic modulators. To overcome this, we use the polarization DoF at Bob and simplify the experimental setup by substituting the Mach-Zehnder configuration (\figurename{~\ref{fig_prin_DDI}~(a)) with a Sagnac interferometer (\figurename{~\ref{fig_prin_DDI}~(b)}). In this way, no active phase stabilization is needed to preserve the state encoded by Alice. Moreover, only one polarization modulator, supplemented with a Faraday rotator and a half-wave plate (HWP), is necessary to encode the same state of polarization in both directions, i.e. clockwise and counter-clockwise.

Our practical implementation is depicted in \figurename{\,\ref{fig_exp_DDI}}. Alice's source starts with a gain-switched DFB laser at 1554.94\,nm (ITU channel C28) triggered at 625\,MHz, which generates light pulses with a duration of 80\,ps. The qubit states are encoded via a set of fiber polarization controllers (PCs) and a birefringent lithium-niobate (LiNbO$_3$) phase modulator (PM) driven by a 3-level pulse generator. Photons enter the PM in the state $(\ket H + \ket V)/\sqrt{2}$ and the effect of the PM is to transform the state into $(\ket H + e^{i\phi}\ket V)/\sqrt{2}$, where $\phi$ is the encoded phase. To compensate the temporal walk-off (around 10 ps) introduced by the birefringence of the modulator, 8 m of polarization-maintaining fiber (PMF) (high-birefringence fiber) is added. The temperature of the birefringent elements (PM and PMF) is actively stabilized to avoid polarization drifts. An additional unitary transformation is performed via a PC placed at the output of the PMF to generate the qubits in the Z or X basis.

\begin{figure*}
\includegraphics[width = 1.6\columnwidth]{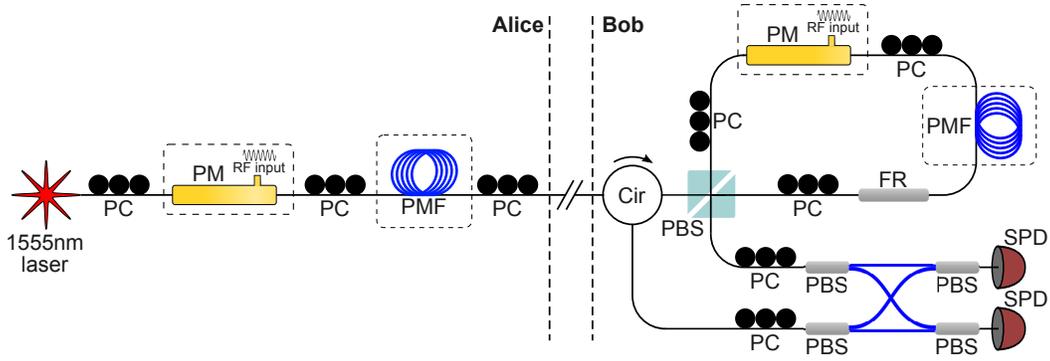}
\caption{\label{fig_exp_DDI}Schematic of the fast DDI-QKD system. PC: polarization controller; PM: phase modulator based on a lithium niobate waveguide; PMF: polarization maintaining fiber; Cir: optical circulator; PBS: polarizing beam splitter; FR: Faraday rotator.}
\end{figure*}

On Bob's side, the polarization qubits of Alice are converted into spatial qubits by a free-space four-port PBS with a polarization rejection superior than 1/1000 for the four arms. As represented in \figurename{\,\ref{fig_exp_DDI}}, at the outputs of the PBS, a Sagnac loop consists of a Faraday rotator, a PM and a PMF, identical to those of Alice. These elements transform both components of Alice's qubit - those that rotate clockwise and counter-clockwise in the Sagnac loop - such that they pass through the PM with the same state of polarization and at the same time. Alice's encoding in polarization has therefore been completely transformed into a spatial encoding. Whilst Bob's qubit is encoded by the PM in the same way that Alice had encoded her own.

To perform the BSM, a PC and a PBS is placed in each output port of the loop. The outputs of the BSM corresponding to $\ket{\Psi^+}$ and $\ket{\Psi^-}$ are delayed by 800\,ps and combined with $\ket{\Phi^+}$ and $\ket{\Phi^-}$, respectively, by means of PBSs. This allows the use of just two detectors for all four Bell states. Bob's setup has 7.1\,dB of attenuation mainly due to the PM ($\approx\,$4\,dB). The photons are detected by two InGaAs/InP negative feedback avalanche diodes operating in the free-running mode and cooled with a free-piston Stirling cooler~\cite{Korzh2014}. The laser, the PMs and the detectors are connected to two field-programmable gate arrays (FPGAs) placed on Alice and Bob's side. A service channel operating on a separate optical fiber is used to synchronize the two parties and to exchange data during the key sifting phase \cite{walenta2014, korzh2015provably}.

For the key exchange protocol, we use the $\mathsf{Z}$ basis to generate the data and the $\mathsf{X}$ basis to estimate the phase error rate. To maximize the key rate, the probability of choosing the $\mathsf{Z}$ basis is 87.5$\%$, both on Alice and Bob's side. To simplify the experimental implementation, Alice uses only three states $\left(\ket{H}; \ket{V}; \ket{+} \right)$~\cite{Fung2006, Tamaki2014}, while Bob uses four states as usual. For a standard BB84 protocol with four states, the QBER in the $\mathsf{X}$ basis is approximated by 
\begin{equation}
p_{\mathsf{X},\tn{err}} \approx \frac{N_{+-} + N_{-+} }{N_{++} + N_{--}  + N_{+-} + N_{-+} },
\end{equation}
with $N_{a,b}$ the number of detections where Alice prepares the state $a$ and Bob's measurement outcome is $b$. For a three state protocol, the phase error rate can be formulated in terms of matched and mismatched bases statistics. In particular, we have:
\begin{multline}
p_{\mathsf{X},\tn{err}} \approx \frac{N_{+-} }{N_{++} + N_{+-} } +\\
\frac{1}{2} \left( \frac{N_{H+} }{N_{H+} + N_{H-} } +\frac{N_{V+} }{N_{V+} + N_{V-} } -1 \right).
\label{eq.QBERx1}
\end{multline}

The SKR is calculated from the error rates in the $\mathsf{Z}$ and $\mathsf{X}$ bases for different transmissions as a function of $\mu$, the mean number of photons per pulse sent by Alice. We make no assumption about the detection efficiency of the detectors as well as the transmission of Bob's setup. The upper bound on the extractable secret key length is given by:
\begin{multline} \label{eq:ell_main}
\ell = \big{\lfloor} s_{\mathsf{Z},1}^{\tn{lb}} (1-\tn{h}_2(\delta_{\mathsf{Z},\tn{ph}}^\tn{ub}))-\tn{leak}_\tn{EC}\\-4\log_2\left(7/\eps_\tn{sec}\right) -\log_2\left( {1}/{\epscorr}\right) \big\rfloor,
\end{multline}
where $s_{\mathsf{Z},1}^{\tn{lb}}$ is the lower bound on the number of single-photon detections in the $\mathsf{Z}$ basis, $\delta_{\mathsf{Z},\tn{ph}}^\tn{ub}$ is the upper bound on the phase error rate, $\tn{leak}_\tn{EC}$ is the number of bits revealed during the error correction step, and $\eps_\tn{sec}$ and $\eps_\tn{cor}$ are the secrecy and correctness parameters, respectively. We fixed the security parameter to $\eps_\tn{qkd} = \eps_\tn{cor} + \eps_\tn{sec} = 4 \times 10^{-9}$, which is similar to those typically used in PtP QKD systems~\cite{korzh2015provably}. Please refer to Appendix~\ref{sec_skr} for more details about equations~\ref{eq.QBERx1} and~\ref{eq:ell_main} as well as their derivation. \\

\begin{figure}
\includegraphics[width = \columnwidth]{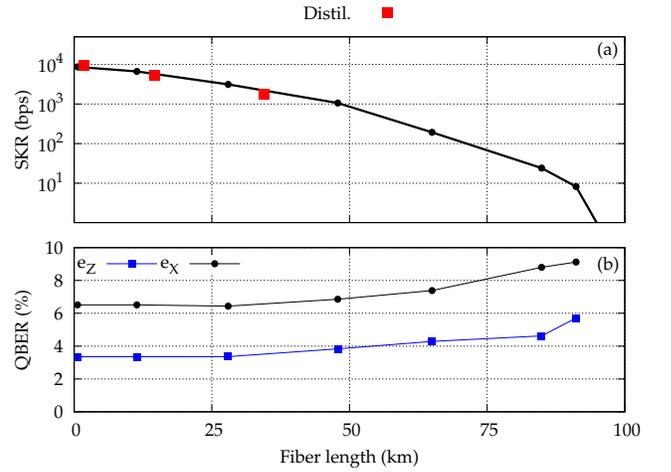}
\caption{\label{fig_SKR}({a}) SKR as a function of the distance. The red squares correspond to complete distillation of a secret key. The black curve corresponds to the SKRs measured without taking into account the finite key statistics. ({b}) QBER in the $\mathsf{Z}$ and $\mathsf{X}$ bases as a function of the distance.}
\end{figure}

\section{Key distillation}

We performed an exchange of secret keys with complete distillation - i.e. including finite key analysis and privacy amplification - at three different distances simulated with a variable attenuator. For every result depicted in \tablename{\,\ref{tab.Distillation}} we optimized the following parameters in order to maximize the SKR: $\mu$, the number of photons per pulse sent by Alice; the dead time of the detectors and the temperature of the detectors. The error correction was carried out using an optimized Cascade algorithm, implemented in C++, which achieved an efficiency of reconciliation of 1.04 for a QBER of 3\%~\cite{Martinez2015}. The efficiency of reconciliation $f_{EC}$ is defined as $f_{EC}=m/(nH(A|B))$, with $m$ the number of bits disclosed during the error correction, $n$ the length of the key before error correction and $H(A|B)$ the conditional entropy between the keys of Alice and Bob before the error correction. In order to reduce the effect of finite-key statistics, the privacy amplification was carried out on a block size of $10^7$\,bits. We obtained a SKR of 1.8~kbps for an attenuation of 6.8~dB corresponding to a distance of 34~km considering 0.2\,dB/km loss.

\begin{table}
\begin{tabular}{C{2.8cm}|C{2.3cm}}
Attenuation [dB]&SKR [kbps]\\
\hline
0.28 & 9.7 \\
2.8 & 5.3 \\
6.8 & 1.8 \\
\end{tabular}
\caption{\label{tab.Distillation}SKR obtained at the output of the system after distillation of block size of $10^7$ bits for different attenuations.}
\end{table}

We also performed exchange of secret keys for additional distances without taking into account the finite key analysis. The corresponding SKRs as a function of the attenuation (converted into fiber distance considering loss of $0.2$\,dB/km) between Alice and Bob are plotted in \figurename{\,\ref{fig_SKR} (\textbf{a})}. We obtained an SKR of 8.2~bps at 91 km. Let us emphasize that these data are obtained from the statistics of actual raw keys. Moreover, the corresponding error rates in the $\mathsf{Z}$ and $\mathsf{X}$ bases are given in \figurename{\,\ref{fig_SKR}} (\textbf{b}). The difference between $e_{x}$ and $e_{z}$ is mainly due to the polarization depend loss of the Sagnac loop elements which introduces a small bias between H and V.

In comparison to a standard BB84 implementation, our system is affected by the significant loss in Bob's device. It could be possible to reduce this loss by at least 2\,dB by changing the PM. Nevertheless, we achieved key exchange over distances up to 91~km without decoy states~\cite{Hwang2003, Lo2005b, Wang2005}, which would not be possible for MDI-QKD. The maximal distance could be significantly improved by adding decoy-state preparation at Alice in order to optimize the bound on the single photon detections. In this configuration we expect to exchange a secret key at a rate of 2~bps over 250\,km of standard single mode fiber. This prediction is based on a simulation which takes into account loss in the apparatus and error rates as measured in our experiment, as well as finite key analysis with a block size of $10^6$ bits.

\section{Conclusion}

We discussed the security of DDI-QKD, which is not equivalent to MDI-QKD in the most general scenario. Nevertheless, we have shown that under very reasonable assumptions, its security can be guaranteed. Although the title ``detector-device-independent'' could be debatable, DDI-QKD offers improved security compared to normal PtP protocols while being easier to implement than MDI-QKD. In particular DDI-QKD requires only single-photon interference, the BSM is 100\% efficient and the performance in the finite-key scenario is similar to PtP QKD.

We implemented a complete high-speed version of the DDI-QKD protocol  clocked at 625~MHz, based on polarization encoded qubits. We distilled secret keys, whilst accounting for finite-key effects, at a rate of 1.8\,kbps for a distance of 34\,km. Furthermore, we achieved a key exchange over 91\,km (without decoy-state preparation).

\section*{Acknowledgements}

We would like to acknowledge Jes\'us Mart\'inez-Mateo for providing the error correction code, and Bing Qi and Marcos Curty for helpful discussions. We thank the Swiss NCCR QSIT and the European EMPIR MIQC2 for financial support. C.~C.~W.~Lim acknowledges support from the Oak Ridge National Laboratory directed research and development program.

\appendix

\section{Estimation of the secret key rate}
\label{sec_skr}

Here, we first present briefly the security analysis of our QKD protocol against a large class of attacks under the assumption that the adversary, Eve, can only forward a qubit or a vacuum state in each use of the quantum channel. Then we show how to estimate the SKR from our experimental raw data. 

In our QKD implementation,~Alice uses a phase-randomized laser source with intensity $\mu$ to prepare her qubits.~In this case, the source generates in each run a vacuum state with probability $\exp(-\mu)$, a single-photon state with probability $\mu\exp(-\mu)$, and a multi-photon state with probability $1-(1+\mu)\exp(-\mu)$. To deal with events that are not single photons, we conservatively assume that (1) multi-photon states are insecure and (2) vacuum states are secure qubit states. The former is due to the fact that Eve can perform photon-number splitting attacks and the latter is due to the fact that vacuum states carry zero information about Alice's bit values; this also applies to Bob. 

Recall that in each run of the implementation, Alice randomly prepares her qubit in one of the three states $\{\ket{0},\ket{1},\ket{+}\}$.~This choice of encoding is known as the \emph{three-state} QKD protocol~\cite{Fung2006}, and it has been recently shown that it is loss-tolerant if mismatched bases statistics are taken into consideration~\cite{Tamaki2014}.~By loss-tolerant, we mean that three-state QKD is resilient against attacks that exploit channel loss and source errors (i.e., encoding flaws).~Interestingly, it has also been shown that the security performance of three-state QKD is similar to BB84 QKD. This means that the fourth qubit state, $\ket{-}$, is redundant. 

Before we state the security bounds for our protocol, it is instructive to spell out the security criteria that we are using. For some small protocol errors, $\epscorr,\esec>0$, we say that our protocol is $\epscorr+\esec$-secure if it is $\epscorr$-correct and $\esec$-secret.~More specifically, let $S_\tn{A}$ and $S_\tn{B}$ be Alice and Bob's output keys,  then the former is satisfied if $\Pr[S_{\rm A}\not=S_{\rm B}] \leq \epscorr$, i.e., the secret keys are identical except with a small probability $\epscorr$. The latter is satisfied if $(1-p_{\rm abort})\|\rho_{\rm AE}-U_{\rm A} \otimes \rho_{\rm E}\|_1/2 \leq \esec$ where $\rho_{\rm AE}$ is the classical-quantum state describing the joint state of $S_{\rm A}$ and $E$,~$U_{\rm A}$ is the uniform mixture of all possible values of $S_{\rm A}$,~and $p_{\rm abort}$ is the probability that the protocol aborts.~Importantly, this secrecy criterion guarantees that the protocol is universally composable: the pair of secret keys can be safely used in any cryptographic task, e.g., for encrypting messages, that requires a perfectly secure key.

To analyze the security of our QKD implementation, we work in a counterfactual scenario where Alice and Bob are using the asymmetric BB84 QKD, i.e., the $\mathsf{Z}$ basis is used for the key and the $\mathsf{X}$ basis is used for parameter estimation.~In this scenario, an upper bound on the extractable secret key length is obtained by using the bound given in Ref.~\citep{Lim2014a}:
\begin{multline} \label{eq:ell}
\ell = \big{\lfloor} s_{\mathsf{Z},1}^{\tn{lb}} (1-\tn{h}_2(\delta_{\mathsf{Z},\tn{ph}}^\tn{ub}))-\tn{leak}_\tn{EC}\\-4\log_2\left(7/\eps_\tn{sec}\right) -\log_2\left( {1}/{\epscorr}\right) \big\rfloor,
\end{multline}
where $s_{\mathsf{Z},1}^{\tn{lb}}$ is the lower bound on the number of single-photon detections in the $\mathsf{Z}$ basis, $\delta_{\mathsf{Z},\tn{ph}}^\tn{ub}$ is the upper bound on the phase error rate, $\tn{leak}_\tn{EC}$ is the number of bits revealed during the error correction step~\footnote{Note that this does not include the information leakage due to error verification.}, and $\tn{h}_2(x)$ is the binary entropy function.~In the following, we show how to compute $s_{\mathsf{Z},1}^{\tn{lb}}$ and $\delta_{\mathsf{Z},\tn{ph}}^\tn{ub}$ using measurement statistics obtained in the actual QKD system.~To start with, we denote by $n_\mathsf{Z}$ and $n_\mathsf{X}$ the total number of detections in the $\mathsf{Z}$ and $\mathsf{X}$ bases, respectively. Then, we have that the total number of detections (in either basis) is a sum of detections conditioned on the number of photons sent by Alice.~For example, for the $\mathsf{Z}$ basis, we have $n_\mathsf{Z}=\sum_k s_{\mathsf{Z},k}$, where $ s_{\mathsf{Z},k}$ is the number of detections conditioned on Alice sending $k$-photon states. Note that since all vacuum states are assumed to be secure qubit states, we may absorb $s_{\mathsf{Z},0}$ into $s_{\mathsf{Z},1}$.

~Let $N_\mathsf{Z}$ be the number of signals with which Alice and Bob choose the $\mathsf{Z}$ basis, then a simple lower bound on $s_{\mathsf{Z},1}$ is obtained by subtracting $G(N_\mathsf{Z})$ the number of multi-photon states sent by Alice from the total number of detections,
\be \label{eq:s1z}
s_{\mathsf{Z},1}^{\tn{lb}}= n_\mathsf{Z} -G(N_\mathsf{Z}),
\ee where
\be
 G(x):=\left\lfloor x(1-(1+\mu)e^{-\mu})+\sqrt{\log(\esec^{-1})x/2} \right\rfloor. 
\ee
Here we used the fact that the photon number distribution follows a Poisson distribution and that maximally $\lfloor N_\mathsf{Z}(1-(1+\mu)e^{-\mu})+\sqrt{\log(\esec^{-1})N_\mathsf{Z}/2} \rfloor$ of the states are multi-photon states; although the latter statement only holds with probability $1-\esec$. Likewise, we have the same bound for the $\mathsf{X}$ basis, 
\be\label{eq:s1x}
s_{\mathsf{X},1}^{\tn{lb}}= n_\mathsf{X} - G(N_\mathsf{X}).
\ee

Next, we need to estimate the number of phase errors in $s_{\mathsf{Z},1}$. In BB84 QKD, this estimation problem is a classical random sampling (without replacement) problem, and one can use the error rate $\delta_\mathsf{X}$ in the $\mathsf{X}$ basis to estimate the phase error rate $\delta_{\mathsf{Z},\tn{ph}}$ in the $\mathsf{Z}$ basis.~However, in three-state QKD, one only has partial observation of the error rate in the $\mathsf{X}$ basis, since only the state $\ket{+}$ is sent in the $\mathsf{X}$ basis.~Recently, it has been shown that the phase error rate can be exactly estimated (in the asymptotic limit) by using the mismatched bases statistics and the partial error rate observed in the $\mathsf{X}$ basis~\cite{Tamaki2014}.~Below, for completeness, we provide an alternative derivation that relates $\delta_\mathsf{X}$ to the above measurement statistics.

Suppose Alice is able to prepare single-photon states and Eve interacts independently and identically with each photon; later we will consider the scenario with weak laser pulses. Furthermore, without loss of generality, we may assume that the quantum channel has perfect transmission, since Bob's basis choice is independent of Eve's attacks and Alice and Bob postselect the measurement statistics.~In this scenario, we may describe Eve's actions using the following transformations:\begin{equation}
\begin{array}{l}
\mathcal{U}_{AE} \ket{H}_A\ket{\phi}_E = \ket{H}_A\ket{\phi_{1}}_E +  \ket{V}_A\ket{\phi_2}_E\\
\mathcal{U}_{AE} \ket{V}_A\ket{\phi}_E = \ket{H}_A\ket{\phi_{3}}_E +\ket{V}_A\ket{\phi_{4}}_E,
\end{array}
\end{equation} where $\ket{\phi_i}$ for $i=1,2,3,4$ are Eve's quantum states (not necessarily normalized). Furthermore, since $\mathcal{U}_{AE}$ is unitary, we have $\bra{\phi_1}\phi_1\rangle +\bra{\phi_2}\phi_2\rangle = 1$, $\bra{\phi_3}\phi_3\rangle +\bra{\phi_4}\phi_4\rangle=1$, and $\bra{\phi_1}\phi_3\rangle +\bra{\phi_2}\phi_4\rangle=0$, $\bra{\phi_3}\phi_1\rangle +\bra{\phi_4}\phi_2\rangle=0$; since the context is now clear, hereafter we will omit the subsystem labels.~Using the above transformations, we thus have
\begin{equation}
\mathcal{U}_{AE} \ket{\pm}\ket{\phi} = \dfrac{\ket{H} \left( \ket{\phi_1} \pm \ket{\phi_3} \right)  + \ket V \left(\ket{\phi_2} \pm \ket{\phi_4} \right)}{\sqrt{2}}.
\end{equation}
Let $a\in \{H,V,+\}$ and $b\in \{H,V,+,-\}$, then the probability that Bob detects $b$ (using the $\mathsf{Z}$ basis) when Alice has sent $a$ (in the $\mathsf{Z}$ basis) follows
\begin{equation}\label{eq_proba}
\begin{array}{l}
p_{\mathsf{Z}\mathsf{Z}}(H|H)= \braket{\phi_1}{\phi_1}\\
p_{\mathsf{Z}\mathsf{Z}}(V|H) = \braket{\phi_2}{\phi_2}\\
p_{\mathsf{Z}\mathsf{Z}}(H|V) = \braket{\phi_3}{\phi_3}\\
p_{\mathsf{Z}\mathsf{Z}}(V|V)= \braket{\phi_4}{\phi_4}.
\end{array}
\end{equation}
From the above, the probabilities for mismatch basis choices are thus given by 
\begin{equation}\label{eq_proba2}
\begin{array}{ll}
p_{\mathsf{X}\mathsf{Z}}(\pm|H) &= \frac{1}{2} \pm \tn{Re}[\braket{\phi_1}{\phi_2}]\\
p_{\mathsf{X}\mathsf{Z}}(\pm|V) &= \frac{1}{2} \pm \tn{Re}[\braket{\phi_3}{\phi_4}]\\
p_{\mathsf{Z}\mathsf{X}}(H|\pm) &=  \frac{\braket{\phi_1}{\phi_1}}{2}+\frac{\braket{\phi_3}{\phi_3}}{2} \pm \tn{Re}[\braket{\phi_1}{\phi_3}]\\
p_{\mathsf{Z}\mathsf{X}}(V|\pm) &= \frac{\braket{\phi_2}{\phi_2}}{2}+\frac{\braket{\phi_4}{\phi_4}}{2} \pm \tn{Re}[\braket{\phi_2}{\phi_4}].\end{array}
\end{equation}
Since $p_{\mathsf{Z}\mathsf{X}}(H|\pm)+p_{\mathsf{Z}\mathsf{X}}(V|\pm)=1$, we have $\tn{Re}[\braket{\phi_1}{\phi_3}]+ \tn{Re}[\braket{\phi_2}{\phi_4}]=0$. Accordingly, we have
\begin{multline}
p_{\mathsf{X}\mathsf{X}}(+|\pm)=\frac{1}{2}\big(1+\tn{Re}[\braket{\phi_1}{\phi_2}]+\tn{Re}[\braket{\phi_3}{\phi_4}]\\ \pm\tn{Re}[\braket{\phi_2}{\phi_3}]\pm\tn{Re}[\braket{\phi_1}{\phi_4}]  \big).
\end{multline}
In the counterfactual BB84 QKD, Alice prepares $\ket{+},\ket{-}$ with uniform probability, and the probability of error in the $\mathsf{X}$ basis is defined as $p_{\mathsf{X},\tn{err}}:=p_{\mathsf{X}\mathsf{X}}(+|-)/2+p_{\mathsf{X}\mathsf{X}}(-|+)/2$.~Using the above equations, we get
\be\label{eq:p_err}
p_{\mathsf{X},\tn{err}}
=\frac{1}{2} \left[p_{\mathsf{X}\mathsf{Z}}(+|H) + p_{\mathsf{X}\mathsf{Z}}(+|V)+2p_{\mathsf{X}\mathsf{X}}(-|+) - 1  \right].
\ee
That is, the probability of observing an error in the $\mathsf{X}$ basis statistics can be exactly estimated by three conditional probabilities:~$p_{\mathsf{X}\mathsf{Z}}(+|H)$, $p_{\mathsf{X}\mathsf{Z}}(+|V)$, and $p_{\mathsf{X}\mathsf{X}}(-|+)$. 

To estimate the phase error rate $\delta_{\mathsf{Z},\tn{ph}}$ in the $\mathsf{Z}$ basis using Eq.~\eqref{eq:p_err}, we have to first estimate $p_{\mathsf{X}\mathsf{Z}}(+|H)$, $p_{\mathsf{X}\mathsf{Z}}(+|V)$, and $p_{\mathsf{X}\mathsf{X}}(-|+)$ from the observed statistics. To start with, let $m(b,a)$ denote the number of detections when Alice sends light pulses prepared in $a\in\{H,V,+\}$ and Bob encodes $b\in\{H,V,+,-\}$, and $m_{\mathsf{X}}(a)=m(+,a)+m(-,a)$.~Then, following the method as described before, we can compute a lower bound on the number of single-photon detections in $m_{\mathsf{X}}(a)$ for any $a$.~For instance, we have $q_{\mathsf{X},1}^{\tn{lb}}(a)=m_{\mathsf{X}}(a)- G(N_{\mathsf{X}(a)})$,
where $N_{\mathsf{X}(a)}$ is the number of instances with which Alice sends $a$ and Bob chooses an encoding in the $\mathsf{X}$ basis.~With that, we can compute upper bounds on the relative frequencies associated with the above conditional probabilities, i.e., we have
\be
f^{\tn{ub}}(\pm|a) = \min\left\{ \frac{1}{2},\frac{m(\pm,a)}{q_{\mathsf{X},1}^{\tn{lb}}(a)}\right\},\quad 
\ee for any $a$. Next, by making use of Hoeffding's inequality, we further get
\be
p_{\mathsf{X}\mathsf{Z}}(+|a) < p^\tn{ub}_{\mathsf{X}\mathsf{Z}}(+|a):=f^{\tn{ub}}(+|a)+ K(q_{\mathsf{X},1}^{\tn{lb}}(a)),
\ee where $K(x):=\sqrt{{2x}/{\log(1/\eps_{\tn{sec}})}}$. Putting everything together, we thus get
\be
p_{\mathsf{X},\tn{err}}^\tn{ub}=\frac{1}{2} \left[p^\tn{ub}_{\mathsf{X}\mathsf{Z}}(+|H) + p^\tn{ub}_{\mathsf{X}\mathsf{Z}}(+|V)+2p^\tn{ub}_{\mathsf{X}\mathsf{X}}(-|+) - 1  \right].
\ee Finally, to compute $\delta_{\mathsf{Z},\tn{ph}}^\tn{ub}$, we use the Hoeffding's inequality again to get
\be
\delta_{\mathsf{Z},\tn{ph}}^\tn{ub}:=p_{\mathsf{X},\tn{err}}^\tn{ub} + K(s_{\mathsf{X},1}^{\tn{lb}}).
\ee

\section{Siphoning attacks on DDI-QKD}
\label{sec_siphoning}

In this section, we present a quantum siphoning attack on BB84 and DDI-QKD. This attack is more powerful than the attack proposed in Ref.~\cite{Qi2015}: it does not require shared randomness between Bob's laboratory and Eve and works even if Bob's input optical mode is restricted to a single spatial-temporal mode.~The central idea of the attack is to exploit the fact that multi-photon states live in the tensor product of single photon subspaces, and linear optical circuits act on each photon independently.~Crucially, these observations suggest that Eve can use multi-photon states to learn about Bob's qubit choices, thereby breaking the security of DDI-QKD.

To illustrate the above idea and to understand the security boundary of DDI-QKD, we start from a conservative scenario whereby the detectors are black boxes and adversarial in nature.~In particular, we assume that the untrusted detectors are controlled by an internal adversary called Fred, who can perform any quantum operation.~Furthermore, we assume that Fred and Eve (who is controlling the quantum channel) are collaborators and they agree on a set of possible actions beforehand.~However, Fred cannot communicate freely with Eve, since Bob's laboratory is secure; however, as we will show below, Eve can communicate freely with Fred.~Also, Fred is restricted to his own device and has no access to Bob's linear optical circuit, e.g., Bob's random basis choices and bit values. 

On Bob's end, we assume that he is able to restrict all input light states to a single optical spatial-temporal mode, where any quantum information is encoded in the polarization DoF.~This assumption is pretty strong as it already allows Bob to rule out a large class of detector side-channel attacks, e.g., time-shifting attacks.~Nevertheless, despite this assumption, we show below that DDI-QKD is insecure if the quantum channel admits multi-photon excitations of the input optical mode.

\begin{figure}[h!]
\includegraphics[width=75mm]{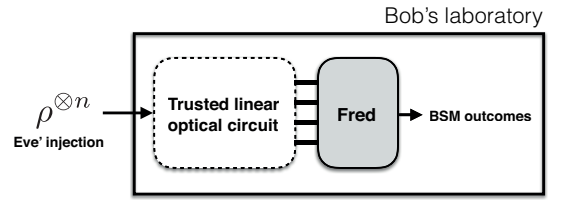}
\caption{\label{Attack Model}With reference to Fig. 1 in the main text, we assume that all the four output ports of Bob's linear optical circuit are given to Fred. In this case, Fred has access to the transformed multi-photon state and can perform any quantum operation on it.~After the operation, he makes a measurement and outputs either one of the four Bell outcomes, $\{\Phi^+$,$\Phi^-$,$\Psi^+$,$\Psi^-\}$, or the inconclusive outcome, $\emptyset $. }
\end{figure}

The quantum siphoning attack is carried out in three phases: (1) the intercept-and-resend phase, (2) the qubit extraction phase, and (3) the siphoning phase. In the first phase, Eve measures Alice's qubit randomly using either the $\mathsf{Z}$ or $\mathsf{X}$ basis, and sends a $n$-photon state to Bob, where each photon is prepared in the equal superposition of the horizontal and vertical polarization states, i.e., $\ket{\chi_{\tn{i}}}_\tn{Eve}=\ket{H}/\sqrt{2}+\ket{V}/\sqrt{2}$.~The number of photons, $n$, is dependent on Eve's basis choice and her measurement outcome: $n$ takes value from the  set $\{n_j\}_{j=1}^{4}$, which corresponds to $\{H,V,+,-\}$.~For example, $n=n_1$ means Eve measures in the $\mathsf{Z}$ basis and obtains $H$ and $n=n_{4}$ means she measures in the $\mathsf{X}$ basis and obtains $-$.~Note that the set $\{n_j\}_{j=1}^{4}$ satisfies $n_j \not= n_k$ for all $j \not=k$ and $n_j \gg 3$ for all $j$.

In the second phase, the $n_j$-photon state passes through Bob's linear optical circuit and each photon is transformed to a four-dimensional quantum state (just before the detectors/Fred):
\be
\ket{\chi^\phi_\tn{f}}_\tn{Eve}=\frac{(1+e^{i\phi})(\ket{1}+\ket{2})	+(1-e^{i\phi})(\ket{3}+\ket{4})}{2\sqrt{2}},
\ee
where $\{\ket{i}\}_i$ is simply the single-photon basis states for the four output ports of Bob's linear optical circuit.~Recall that Bob's qubit choice is denoted by $\phi\in\{0,\pi,\pi/2,3\pi/2\}$.~In fact, we can further simplify the above equation to reflect an effective qubit state by using the transformations: $\ket{\tilde{0}}=(\ket{1}+\ket{2})/\sqrt{2}$ and $\ket{\tilde{1}}=(\ket{3}+\ket{4})/\sqrt{2}$, $\ket{\tilde{\pm}}=(\ket{\tilde{0}}\pm\ket{\tilde{1}})/\sqrt{2}$, giving 
\be
\ket{\chi^\phi_\tn{f}}_\tn{Eve}=\frac{(1+e^{i\phi})}{2}\ket{\tilde{0}}+\frac{(1-e^{i\phi})}{2}\ket{\tilde{1}}.	
\ee
Indeed, we see that when the input is a single-photon state, the resulting output states are the BB84 qubit states (up to local rotations), where one basis is given by
\be
\ket{\chi^0_\tn{f}}_\tn{Eve} =\ket{\tilde{0}}, \quad
\ket{\chi^\pi_\tn{f}}_\tn{Eve}=\ket{\tilde{1}},\ee and the other basis given by
\be
\ket{\chi^{\pi/2}_\tn{f}}_\tn{Eve} =\frac{\ket{\tilde{+}}+i\ket{\tilde{-}}}{\sqrt{2}},\quad \ket{\chi^{3\pi/2}_\tn{f}}_\tn{Eve}=\frac{\ket{\tilde{+}}-i\ket{\tilde{-}}}{\sqrt{2}}.
\ee
When the input state is a $n_j$-photon state, the output is $n_j$ copies of Bob's qubit; because Fred has access to all the four output ports. This means that Fred can first measure the photon number of the output state and learn about Eve's basis choice and her measurement outcome. Then, Fred can determine $\phi$, i.e., Bob's qubit choice by performing the optimal unambiguous state discrimination (USD) measurement for $n_j$ photons. In particular, the regime in which Fred can unambiguously learn about $\phi$ starts from $n=3$, with a success probability of $1/2$. That is, there exists an USD measurement which Fred can perform to extract $\phi$ from 3 copies of $\ket{\chi^\phi_\tn{f}}_\tn{Eve}$ with probability at least $1/2$. In the general case where $n\geq3$, it can be shown that the probability of success for a $n_j$-photon injection is lower bounded by the smallest eigenvalue of the following $2\times2$ block matrix~\cite{Sun2002}:
\be
p_\tn{succ}(n_j)\geq \lambda_\tn{min} \left( \begin{bmatrix}
     \mathbbm{I}_2 & C(n_j)\\
    C(n_j) & \mathbbm{I}_2
\end{bmatrix} \right),
\ee where
\[ \mathbbm{I}_2:= \begin{bmatrix} 1 & 0  \\ 0 &1   \end{bmatrix},\quad C({n_j}):= \begin{bmatrix} \left(\frac{1}{\sqrt{2}}\right)^{n_j}&\left(\frac{1}{\sqrt{2}}\right)^{n_j} \\\left(\frac{1}{\sqrt{2}}\right)^{n_j} &\left(\frac{-1}{\sqrt{2}}\right)^{n_j}  \end{bmatrix}.\] 
Indeed, in the case of ${n_j}=3$, we see that $p_\tn{succ}(3)\geq 1/2 $. Since Eve can inject an arbitrary number of photons, we have to assume the limiting case and take $p_\tn{succ}({n_j}) \approx 1$ with $n_j \gg 3$ for any $j$.

In the final phase, Fred first compares Eve and Bob's basis choices.~If they are the same, he simply outputs a Bell state that is consistent with their bit values and basis choice, otherwise, he announces the measurement as inconclusive.~In particular, Fred uses Tab.~\ref{tab_truthtable} to determine the BSM outcome. For example, if Eve's outcome is $V$ and Bob's qubit choice is $H$, then Fred outputs $\Psi^+$ or $\Psi^-$ with probability 1/2 each.~Note that this attack works whenever the quantum channel loss is $\geq 1/2$. 

The above quantum siphoning attack works even if Bob limits the input light state to a single optical spatial-temporal mode. Our proposed attack is essentially an entanglement-breaking operation, since it requires Eve to perform an intercept-and-resend attack. More crucially, it should be noted that this attack works as long as Eve is able to send multi-photon states to Bob.~This problem is reminiscent of the security problem faced by the bi-directional~``plug \& play" QKD system, where Eve can apply Trojan horse attacks to learn about Alice's bit values.~Such Trojan horse attacks could be mitigated by employing countermeasures like those proposed in Ref.~\cite{Zhao2008b}. To conclude, our attack shows that DDI-QKD is not equivalent to MDI-QKD, despite their conceptual similarities, and additional assumptions are necessary to guarantee the security of DDI-QKD as discussed in Sec.~\ref{sec.security}.

\bibliography{bib_ddiqkd}
\end{document}